\documentstyle[12pt,epsfig,axodraw]{article}
\def\be{\begin{equation}}
\def\ee{\end{equation}}
\def\bc{\begin{center}}
\def\ec{\end{center}}
\def\bea{\begin{eqnarray}}
\def\eea{\end{eqnarray}}

\def\nn{\nonumber}
\def\am{{\alpha_{\mu}}}
\def\amc{{\alpha_{\mu^c}}}

\def\auxm{{F^M}}
\def\auxmc{{F^{M^c}}}
\def\auxz{{F^Z}}

\def\cl{{\cal L}}

\def\dmu{\partial^\mu}
\def\dmd{\partial_\mu}

\def\fz{{\psi_z}}

\def\gold{{\psi}_z}
\def\goldb{\overline{\psi}_z}

\def\lsq{{\Lambda^2}}
\def\mgau{M_{\lambda}}

\def\muc{{\mu^c}}
\def\ov{\overline}
\def\sml{{\tilde{\mu}}}
\def\smr{{\overline{\tilde{\mu}^c}}}

\def\st{{\sin \theta}}
\def\ct{{\cos \theta}}
\def\simlt{\stackrel{<}{{}_\sim}}

\def\tm{\tilde{\mu}}
\def\tmc{{\tilde{\mu^c}}}
\def\tmu{{\tilde{\mu}_1}}
\def\tmd{{\tilde{\mu}_2}}

\catcode`@=11
\def\marginnote#1{}
\newcount\hour
\newcount\minute
\newtoks\amorpm
\hour=\time\divide\hour by60
\minute=\time{\multiply\hour by60 \global\advance\minute by-\hour}
\edef\standardtime{{\ifnum\hour<12 \global\amorpm={am}%
        \else\global\amorpm={pm}\advance\hour by-12 \fi
        \ifnum\hour=0 \hour=12 \fi
        \number\hour:\ifnum\minute<10 0\fi\number\minute\the\amorpm}}
\edef\militarytime{\number\hour:\ifnum\minute<10 0\fi\number\minute}
\def\draftlabel#1{{\@bsphack\if@filesw {\let\thepage\relax
   \xdef\@gtempa{\write\@auxout{\string
      \newlabel{#1}{{\@currentlabel}{\thepage}}}}}\@gtempa
   \if@nobreak \ifvmode\nobreak\fi\fi\fi\@esphack}
        \gdef\@eqnlabel{#1}}
\def\@eqnlabel{}
\def\@vacuum{}
\def\draftmarginnote#1{\marginpar{\raggedright\scriptsize\tt#1}}
\def\draft{\oddsidemargin 0.0truein
        \def\@oddfoot{\sl preliminary draft \hfil
        \rm\thepage\hfil\sl\today\quad\militarytime}
        \let\@evenfoot\@oddfoot \overfullrule 3pt
        \let\label=\draftlabel
        \let\marginnote=\draftmarginnote
   \def\@eqnnum{(\theequation)\rlap{\kern\marginparsep\tt\@eqnlabel}%
\global\let\@eqnlabel\@vacuum}  }
\catcode`@=12
\begin{document}
\begin{titlepage}
\vspace*{-1cm}
\phantom{hep-ph/9904367} 
\hfill{DFPD-99/TH/12}
\vskip 2.0cm
\begin{center}
{\Large\bf On the muon anomalous magnetic moment \\ 
in models with a superlight gravitino \footnote{Work supported 
in part by the European Commission TMR Programme ERBFMRX-CT96-0045.}}
\end{center}
\vskip 1.5  cm
\begin{center}
{\large 
Andrea Brignole\footnote{E-mail address: 
 brignole@padova.infn.it},
Elena Perazzi\footnote{E-mail address: 
perazzi@padova.infn.it}
and 
Fabio Zwirner\footnote{E-mail address: 
zwirner@padova.infn.it}}
\\
\vskip .5cm
Istituto Nazionale di Fisica Nucleare, Sezione di Padova, 
\\
Dipartimento di Fisica `G.~Galilei', Universit\`a di Padova, 
\\
Via Marzolo n.8, I-35131 Padua, Italy
\end{center}
\vskip 3.0cm
\begin{abstract}
\noindent
We perform a general analysis of the contributions to the muon 
anomalous magnetic moment in models with a superlight gravitino.
We discuss the interpretation and the phenomenological implications
of the results. We find that present constraints on the model 
parameters are comparable and complementary to the ones coming 
from collider searches and from perturbative unitarity. However, 
the Brookhaven E821 experiment will probe large unexplored regions of
parameter space.
\end{abstract}
\end{titlepage}
\setcounter{footnote}{0}
\vskip2truecm
\bc
{\bf 1. Introduction}
\ec

The anomalous magnetic moment of the muon is both one of the best 
measured quantities in particle physics and one of the most 
accurately known predictions of the Standard Model (SM) of strong 
and electroweak interactions. In terms of $a_{\mu} \equiv (g_{\mu} 
- 2)/2$, the present experimental and theoretical situation can 
be summarized \cite{cm} as follows:
\be
\label{al}
\begin{array}{ccrcrcc}
a_{\mu}^{ex} & = & (116592350 & \pm & 730) & \times & 10^{-11} \, ,
\\
a_{\mu}^{SM} & = & (116591596 & \pm &  67) & \times & 10^{-11} \, .
\end{array}
\ee
As a consequence, powerful constraints on extensions of the SM can 
be extracted from the comparison of eq.~(\ref{al}) with possible 
non-standard contributions, $\delta a_{\mu}$. The $95 \%$ 
confidence level bound corresponding to the present data 
is\footnote{More conservative estimates of the present theoretical 
error associated with the hadronic contribution can enlarge the 
total theoretical error by up to a factor of two. Given the present 
experimental precision, this would not affect significantly the 
bounds of eq.~(\ref{limit}), and improvements are expected in the 
future.}
\be
\label{limit}
- 0.7 \times 10^{-8} < \delta a_{\mu} < 2.2 \times 10^{-8} \, ,
\ee
and the E821 experiment at Brookhaven \cite{e821} is expected to 
reduce further the experimental error by roughly a factor of 20.

The most popular and theoretically motivated extensions of
the SM are those incorporating low-energy supersymmetry.
The study of $a_{\mu}$ in supersymmetric models has a long 
history. Already in the early days of supersymmetry
it was realized \cite{fr} that $a_{\mu}=0$ in the limit of
unbroken supersymmetry, both in the global and in the 
local \cite{exsugra} case. In the case of broken supersymmetry, 
the first estimates of $\delta
a_{\mu}$ from loop diagrams involving supersymmetric 
particles were given in \cite{fayet}. By now, the full 
one-loop contribution to $a_{\mu}$ in the Minimal 
Supersymmetric extension of the Standard Model (MSSM) 
is available, for an arbitrary supersymmetric particle 
spectrum \cite{bm,gm2mssm}: the relevant one-loop diagrams 
involve smuon-smuon-neutralino and chargino-chargino-sneutrino 
loops. Thanks to the fast decoupling properties of the soft 
supersymmetry-breaking mass terms, these contributions are 
comfortably within the bounds of eq.~(\ref{limit}) for typical 
values of the MSSM parameters allowed by the direct searches 
for supersymmetric particles. The only situation in which 
$\delta a_{\mu}$ can saturate the bounds of eq.~(\ref{limit})
is when the masses of supersymmetric particles are close to 
their present lower bounds and $\tan \beta = v_2/v_1$ is very
large (so that the muon Yukawa coupling is considerably enhanced).

In supersymmetric models with a light gravitino, the effective 
low-energy theory does not contain only the MSSM states, but 
also the gravitino and its superpartners, and we must compute 
the additional contributions $\delta a_{\mu}$ coming from the 
one-loop diagrams where these particles (as well as other MSSM 
particles) are exchanged on the internal lines. Most of the 
existing calculations of these effects were performed in the 
framework of supergravity, and the discussion \cite{sugra,mo,paco,lln} 
was mainly focused (with some controversies) on the finiteness of 
the broken supergravity contributions. The most complete study 
available so far is the one of ref.~\cite{paco}, which considered 
a general spontaneously broken supergravity, and computed the 
one-loop diagrams involving virtual particles from the matter, 
gauge and gravitational supermultiplets. However, in the case of a
light gravitino (the only phenomenologically relevant one for this 
type of study) we can work directly in the globally supersymmetric 
limit, keeping only the goldstino and its spin--0 superpartners 
(sgoldstinos) as the relevant degrees of freedom from the 
supersymmetry-breaking sector \cite{equiv}. Indeed, this method 
has already been partially applied in \cite{bm}, to compute the 
contribution to $\delta a_{\mu}$ of the goldstino-smuon-smuon 
loops. In this paper, we shall adopt such a method to perform a 
general analysis of the contributions to $\delta a_{\mu}$ from the 
supersymmetry-breaking sector of the theory. 

The paper is organized as follows. In sect.~2, we introduce the
model and derive the relevant interaction terms. In sect.~3, we 
present the general expressions for our results. In sect.~4, 
we discuss their interpretation, with emphasis on possible 
divergent contributions. In sect.~5, we specialize our formulae
to some phenomenologically interesting limits and discuss their
implications. Whenever possible, we compare with the previous 
literature, in particular with the supergravity computation of
ref.~\cite{paco}.

\vspace{1cm}
\bc
{\bf 2. The model}
\ec

We will compute the anomalous magnetic moment of the muon in a general
$N=1$ globally supersymmetric model containing a $U(1)$ 
gauge vector superfield, $V\equiv(\lambda,A_{\mu},D)$, and three
chiral superfields, $M\equiv(\tm,\mu,\auxm)$, $M^c\equiv(\tmc,\muc,
\auxmc)$, $Z \equiv (z,\fz,\auxz)$, with the following charge
assignments: $Q(M)=-1$, $Q(M^c)=+1$, $Q(Z)=0$. Despite its 
simple field content, the model should reproduce all the relevant aspects
of the realistic case. In particular, the $U(1)$ will be associated 
with the exact gauge symmetry of supersymmetric QED, the $M$ and $M^c$
multiplets will contain the degrees of freedom of the left-handed
muon and anti-muon, respectively, and the $Z$ multiplet 
will contain the goldstino. The most general effective 
Lagrangian with the above field content is determined, up to 
higher-derivative terms, by a superpotential $w(Z,M,M^c)$, 
a gauge kinetic function $f(Z,M,M^c)$ and a K\"ahler potential 
$K(Z,\ov{Z},M,\ov{M},M^c,\ov{M^c})$. 
We can parametrize such functions as follows:
\be
\label{w}
w = F Z+ {\sigma  \over 6} Z^3 
+ m M M^c + \rho Z  M  M^c + \ldots \, ,
\ee 
\be
\label{f}
f = {1 \over e^2} \left( 1 + \eta {2  \, Z \over \Lambda}
+ \gamma_f { M M^c \over \Lambda^2} + \ldots \right) \, ,
\ee
$$
K =  
|Z|^2 + |M|^2 + |M^c|^2 - \alpha_z { |Z|^4 \over 4 \lsq }
$$
\be
- \am {|Z|^2 |M|^2 \over \lsq } - \amc {|Z|^2 |M^c|^2 \over \lsq }
- \gamma_K {  \left( \ov{Z}^2 M  M^c + {\rm h.c.} \right) \over 2 \lsq }
+ \ldots \, .
\label{k}
\ee
In the above expressions, all the parameters have been taken to be 
real for simplicity, and the dots stand for terms that either do not 
play any r\^ole in the calculation of $a_{\mu}$ or can be eliminated 
by (non-linear) analytic field redefinitions\footnote{The above 
parametrization corresponds essentially to choosing normal 
coordinates \cite{grk}, and simplifies our diagrammatic computations. 
Although reparametrization covariance is lost in the intermediate
steps, general arguments guarantee that the final physical results 
are the same. We have explicitly checked this, starting from
a fully general parametrization.}. 
For example, a possible cubic term in $K$ of the form $(\beta_K 
\ov{Z} M M^c/\Lambda + {\rm h.c.})$
can be eliminated by the analytic field redefinition $Z \rightarrow
(Z - \beta_K M M^c / \Lambda)$, accompanied by the parameter redefinitions
$(m - \beta_K F / \Lambda) \rightarrow m$ and $(\gamma_f - 2 \eta \beta_K)
\rightarrow \gamma_f$. Other field redefinitions have been used to 
eliminate other low-order terms and to shift $\langle z \rangle$ to
zero. 
Taking all this into account, the above expressions 
of $w$, $f$ and $K$ are the most general ones compatible with a classical 
vacuum with broken supersymmetry and unbroken CP and $U(1)$. Indeed, 
the parametrization is still slightly redundant, since one of the 
dimensionless parameters, e.g. $\eta$, can be re-absorbed in the 
definition of the mass scale $\Lambda$ characterizing the 
non-renormalizable operators, but we stick to it to have a clearer 
physical interpretation of our results.  

By standard techniques, it is easy to check that there is a local 
minimum of the classical potential where supersymmetry is 
spontaneously broken, with vacuum energy $\langle V \rangle = F^2$, 
and the gauge symmetry remains unbroken, with the gauge coupling 
constant given by $\langle ({\rm Re} \, f)^{-1} \rangle = e^2$.
Notice that the K\"ahler metric is canonical at the minimum, so that the 
component fields of the chiral superfields are automatically normalized. 
In the fermion sector of the model, $\psi_z$ is the massless goldstino, 
whereas the muon has a Dirac mass $m_{\mu} = m$, and the photino has 
a Majorana mass $\mgau=\eta F/\Lambda$. In the scalar sector of the 
model, the real and imaginary parts of the complex sgoldstino, $z 
\equiv (S + i P)/\sqrt{2}$, have squared masses:
\be
m_S^2 = \alpha_z {F^2 \over \lsq} + \sigma F \, ,
\;\;\;\;\;
m_P^2 = \alpha_z { F^2 \over \lsq} - \sigma F \, ,
\ee 
and the smuons have the following squared mass matrix:
\be
{\cal M}_0^2 = 
\left( \begin{array}{cc}
\tilde{m}_{\mu}^2 + m_{\mu}^2 & \delta m^2 \\
\delta m^2 & \tilde{m}_{\mu^c}^2 + m_{\mu}^2
\end{array} \right) \, ,
\ee 
where
\be
\tilde{m}_{\mu}^2 = \am {F^2 \over \lsq} \, ,
\;\;\;\;\;
\tilde{m}_{\mu^c}^2 = \amc {F^2 \over \lsq} \, ,
\;\;\;\;\;
\delta m^2 = \rho \, F \, .
\ee
We denote the smuon mass eigenstates by $\tmu \equiv \ct \sml + \st 
\smr$, $\tmd \equiv - \st \sml + \ct \smr$, and the corresponding 
mass eigenvalues by $m_1^2$ and $m_2^2$, respectively. We also recall
that
\be
\label{mixing}
\sin 2 \theta = {2 \delta m^2 \over m_1^2 - m_2^2} \, .
\ee

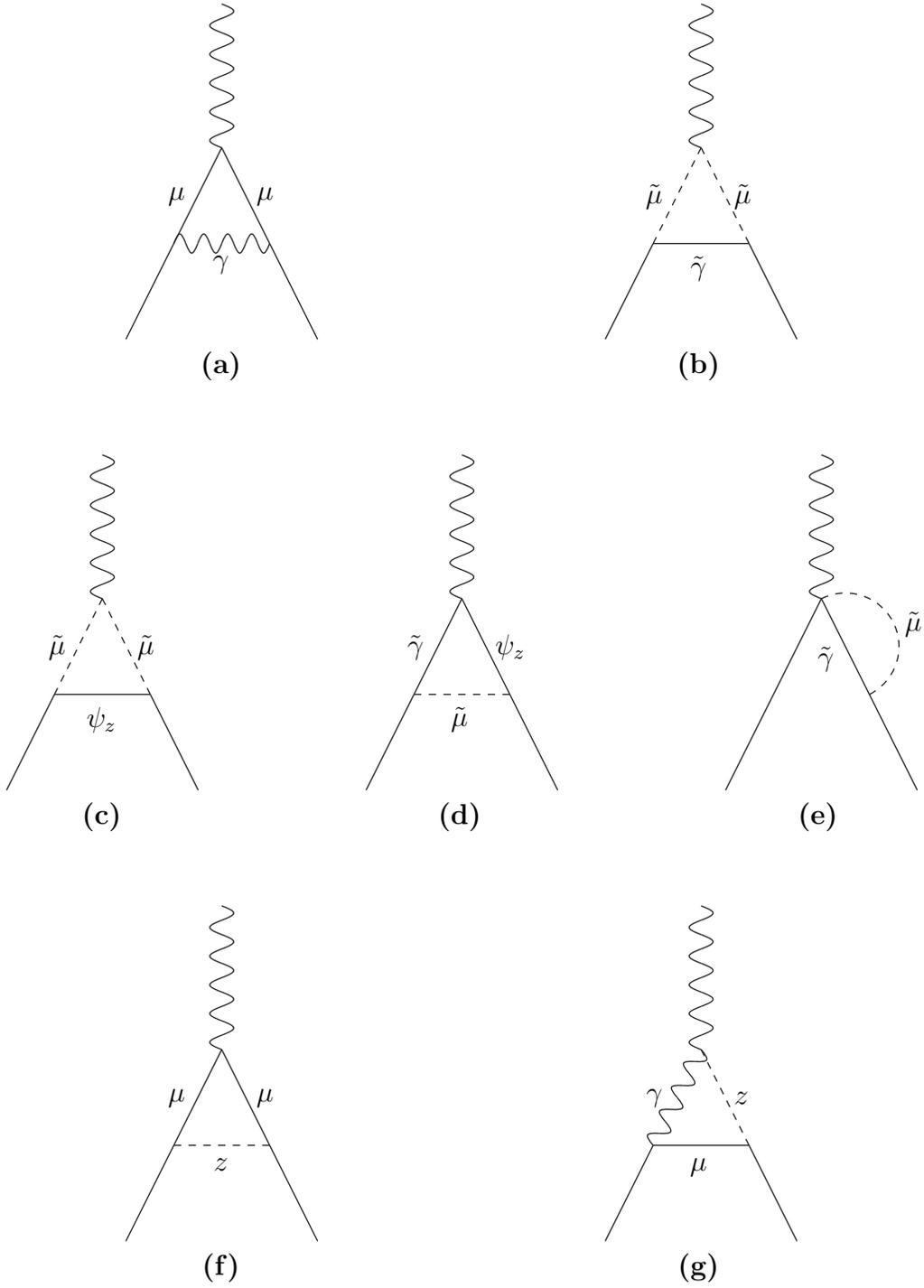
\begin{figure}[htbp]
\begin{center}
\begin{picture}(450,200)(-20,25)
\Photon(100,180)(100,120){5}{5}
\Line(80,80)(100,120)
\Line(60,40)(80,80)
\Line(100,120)(120,80)
\Line(120,80)(140,40)
\Photon(80,80)(120,80){4}{4}
\Text(100,75)[t]{$\gamma$}
\Text(85,100)[r]{$\mu$}
\Text(115,100)[l]{$\mu$}
\Text(100,35)[t]{{\bf(a)}}
\Photon(300,180)(300,120){5}{5}
\DashLine(280,80)(300,120){3}
\Line(260,40)(280,80)
\DashLine(300,120)(320,80){3}
\Line(320,80)(340,40)
\Line(280,80)(320,80)
\Text(285,100)[r]{$\tilde{\mu}$}
\Text(315,100)[l]{$\tilde{\mu}$}
\Text(300,75)[t]{$\tilde{\gamma}$}
\Text(300,35)[t]{{\bf(b)}}
\end{picture}
\end{center}
\begin{center}
\begin{picture}(450,200)(30,0)
\Photon(100,180)(100,120){5}{5}
\DashLine(80,80)(100,120){3}
\Line(60,40)(80,80)
\DashLine(100,120)(120,80){3}
\Line(120,80)(140,40)
\Line(80,80)(120,80)
\Text(85,100)[r]{$\tilde{\mu}$}
\Text(115,100)[l]{$\tilde{\mu}$}
\Text(100,75)[t]{$\psi_z$}
\Text(100,35)[t]{{\bf(c)}}
\Photon(250,180)(250,120){5}{5}
\Line(230,80)(250,120)
\Line(210,40)(230,80)
\Line(250,120)(270,80)
\Line(270,80)(290,40)
\DashLine(230,80)(270,80){3}
\Text(235,100)[r]{$\tilde{\gamma}$}
\Text(265,100)[l]{$\psi_{z}$}
\Text(250,75)[t]{$\tilde{\mu}$}
\Text(250,35)[t]{{\bf(d)}}
\Photon(400,180)(400,120){5}{5}
\Line(360,40)(400,120)
\Line(400,120)(440,40)
\DashCArc(410,100)(22.36,-63.435,116.565){3}
\Text(400,95)[l]{$\tilde{\gamma}$}
\Text(440,115)[t]{$\tilde{\mu}$}
\Text(400,35)[t]{{\bf(e)}}
\end{picture}
\end{center}
\begin{center}
\begin{picture}(450,200)(-20,-25)
\Photon(100,180)(100,120){5}{5}
\Line(80,80)(100,120)
\Line(60,40)(80,80)
\Line(100,120)(120,80)
\Line(120,80)(140,40)
\DashLine(80,80)(120,80){3}
\Text(100,75)[t]{$z$}
\Text(85,100)[r]{$\mu$}
\Text(115,100)[l]{$\mu$}
\Text(100,35)[t]{{\bf(f)}}
\Photon(300,180)(300,120){5}{5}
\Photon(280,80)(300,120){4}{4}
\Line(260,40)(280,80)
\DashLine(300,120)(320,80){3}
\Line(320,80)(340,40)
\Line(280,80)(320,80)
\Text(285,100)[r]{$\gamma$}
\Text(315,100)[l]{$z$}
\Text(300,75)[t]{$\mu$}
\Text(300,35)[t]{{\bf(g)}}
\end{picture}
\end{center}
\vspace{-1.5cm}
\caption{\em The different classes of one-loop diagrams 
contributing to $a_{\mu}$.} 
\label{fig1}
\end{figure}

In the model defined by eqs.~(\ref{w})--(\ref{k}), the different 
classes of one-loop Feynman diagrams that may contribute to $a_{\mu}$ 
are displayed in fig.~\ref{fig1}.
The Feynman rules needed to compute all these diagrams are 
summarized in the following Lagrangian
\bea
\cl & = & 
- {1 \over 4} F_{\mu \nu} F^{\mu \nu} 
+ i \ov{\lambda} \ov{\sigma}^{\mu} \dmd \lambda 
- {\mgau \over 2} \left( \lambda \lambda + \ov{\lambda} 
\, \ov{\lambda} \right) 
+ {1 \over 2} (\dmu S)(\dmd S) - {1 \over 2} m_S^2 S^2
 \nn \\ & & \nn \\ 
& & 
+ {1 \over 2} (\dmu P)(\dmd P) - {1 \over 2} m_P^2 P^2
+ \ov{(D^{\mu} \tmu)} (D_{\mu} \tmu) - m_1^2 |\tmu|^2 
+ \ov{(D^{\mu} \tmd)} (D_{\mu} \tmd) - m_2^2 |\tmd|^2 
\nn \\ & & \nn \\
& &
+ i \goldb \ov{\sigma}^{\mu} \dmd \gold
+ i \ov{\mu} \, \ov{\sigma}^{\mu} D_{\mu} \mu
+ i \ov{\mu^c} \, \ov{\sigma}^{\mu} D_{\mu} \mu^c
- m_{\mu} \left( \mu \mu^c + \ov{\mu} \, \ov{\mu^c} \right) 
\nn \\ & & \nn \\ 
& &
- {1 \over \sqrt{2}} {\mgau \over F} \left(
i \gold \sigma^{\mu \nu} \lambda F_{\mu \nu}
+ {\rm h.c.} \right) 
- {1 \over 2 \sqrt{2}} {\mgau \over F} S \, F^{\mu \nu} F_{\mu \nu} 
+ {1 \over 4 \sqrt{2}} {\mgau \over F} P \,  \epsilon^{\mu \nu \rho 
\sigma} F_{\mu \nu} F_{\rho \sigma}
\nn \\ & & \nn \\ 
& &
-{1 \over \sqrt{2}} \left(  {\delta m^2 \over F}
+ \gamma_K {F \over \lsq} \right) \left( S \mu 
\mu^c + {\rm h.c.} \right)
-{1 \over \sqrt{2}} \left(  {\delta m^2 \over F}
- \gamma_K {F \over \lsq} \right) \left( i \, P 
\mu \mu^c + {\rm h.c.} \right)
\nn \\ & & \nn \\
&  & 
- {1 \over F} \left\{ \left[ (m_1^2 - m_{\mu}^2) \cos \theta 
\ov{\tmu} - (m_2^2 - m_{\mu}^2) \sin \theta \ov{\tmd} \right] 
\mu \gold + {\rm h.c.} \right\}
\nn \\ & & \nn \\
&  & 
- {1 \over F} \left\{ \left[ (m_1^2 - m_{\mu}^2) \sin \theta 
\tmu + (m_2^2 - m_{\mu}^2) \cos \theta \tmd \right] \mu^c 
\gold + {\rm h.c.} \right\} 
\nn \\ & & \nn \\
& & 
+ e \sqrt{2} \left[ (\cos \theta \ov{\tmu} - \sin
\theta \ov{\tmd}) \mu \lambda - (\sin \theta \tmu 
+ \cos \theta \tmd) \mu^c \lambda + {\rm h.c.} \right] 
\nn \\ & & \nn \\
& & 
- \gamma_f {1 \over 2 \sqrt{2} \lsq} \left( i \,
\tilde{\mu} \mu^c \sigma^{\mu \nu} \lambda F_{\mu \nu}
+ i \, \tilde{\mu^c} \mu \sigma^{\mu \nu} \lambda F_{\mu \nu}
+ {\rm h.c.} \right) + \ldots \, ,
\label{leff}
\eea
where we used two-component spinor notation in the conventions
of ref.~\cite{bfz3} (with canonically normalized photon and 
photino fields, and the field redefinition $\lambda \rightarrow
i  \lambda$), and the dots stand for the gauge-fixing term and
other terms that are not relevant for the following calculations. 
In eq.~(\ref{leff}), $\epsilon^{0123} = - \epsilon_{0123} = - 1$, 
and the covariant derivatives are defined according to the following 
sign convention: $D_{\mu} \equiv \dmd + i e Q A_{\mu}$.

To conclude the description of the model, we observe that almost all
the parameters controlling the interactions of eq.~(\ref{leff}) are
related with the supersymmetry-breaking scale, $\sqrt{F}$, and the
particle spectrum (masses and mixing angles): the only two additional
parameters are $\gamma_f$ and $\gamma_K$.

\vspace{1cm}
\bc
{\bf 3. General results}
\ec

We now move to the computation of the different diagrams depicted 
in fig.~1 and to the presentation of the results, without any 
assumption on the model parameters.

The diagrams in figs.~1a and 1b are nothing else than the 
muon-muon-photon and smuon-smuon-photino diagrams of 
supersymmetric QED, whose contributions are well-known. 
Fig.~1a gives the one-loop QED result \cite{schwinger}:
\be
\label{qed}
a_{\mu}^{(a)} = {\alpha \over 2 \pi} \, .
\ee
Its supersymmetric counterpart, fig.~1b, gives
\bea
\delta a_{\mu}^{(b)} & = & - {\alpha \over 2 \pi} \, m_{\mu}^2 \,
\left[ \int_0^1 dx {x^2 (1-x) \over
m_1^2 x + \mgau^2 (1-x) - m_{\mu}^2 x (1-x)}
+ (1 \to 2) \right] \nn \\ & & \nn \\ & & + 
{\alpha \over \pi}  {m_{\mu}  \, \delta m^2 \, \mgau 
\over m_1^2 - m_2^2} \left[ \int_0^1 dx {x (1-x) 
\over m_1^2 x + \mgau^2 (1-x) - m_{\mu}^2 x (1-x)}
- (1 \to 2) \right]  \, .
\label{amsqed}
\eea

Most of the remaining diagrams of fig.~\ref{fig1} involve 
goldstino or sgoldstino exchanges. Some of them are finite,
others lead to potential divergences and require a 
regularization. We will use two alternative regularizations,
appropriate for supersymmetry: either dimensional reduction
in D-dimensional momentum space, or a na\"{\i}ve cutoff 
$\Lambda_{UV}$ in 4-dimensional momentum space. To keep 
track of the regularization dependence, we will use the 
abbreviations:
\be
\label{deltauv}
\Delta_{UV} = \left\{ \begin{array}{ll}
{2 \over 4 - D } - {\gamma_E} + \log 4\pi &
({\rm dim. \; reduction \,} ) \\
\log {\Lambda_{UV}^2 \over \mu^2} - 1 & 
({\rm mom. \; cutoff \,} ) \end{array} \right.
\, ,
\ee
\be
\label{damureg}
{\hat{\Delta}}  = \left\{ \begin{array}{ll}
1 & ({\rm dim. \; reduction \,} ) \\
0 & ({\rm mom. \; cutoff \,} ) \end{array} \right.
\, ,
\ee
where $\gamma_E$ is the Euler-Mascheroni constant and
$\mu$ is the renormalization scale.

The diagrams of class (c), containing smuon-smuon-goldstino 
loops, give a finite result:
\be
\label{ama}
\delta a_{\mu}^{(c)} = - {m_{\mu}^2 \over 16 \pi^2 F^2} 
\left[  \int_0^1 dx {(m_1^2 - m_{\mu}^2)^2 x (1-x) \over
m_1^2 - m_{\mu}^2 (1-x)} + (1 \to 2) \right] \, .
\ee

The diagrams of class (d), containing photino-smuon-goldstino 
loops, are superficially divergent. However, the overall result
is again finite:
\bea
\label{amff}
\delta a_{\mu}^{(d)} &  = &   
- {m_{\mu} \delta m^2 \mgau \over 8 \pi^2 F^2} 
\left[ {\hat{\Delta}}
+ {  2 m_{\mu}^2 \over m_1^2-m_2^2}
\int_0^1 \! \! \! \! dx \int_0^{1-x} \! \! \! \! dy 
\left( 
{ (m_1^2 - m_{\mu}^2) x^2
\over  m_1^2 x +  \mgau^2 y - m_{\mu}^2 x(1-x)}
- (1 \to 2) 
\right) \right]
\nonumber
\\
& &
+ {m_{\mu}^2 \mgau^2 \over 8 \pi^2 F^2} 
\int_0^1  \! \! \! \! dx  \int_0^{1-x} \! \! \! \! dy
\left( 
{ (m_1^2 - m_{\mu}^2) x
\over  m_1^2 x + \mgau^2 y - m_{\mu}^2 x(1-x)}
+ (1 \to 2) 
\right)  \, ,
\eea
where ${\hat{\Delta}}$ was defined in eq.~(\ref{damureg}). Notice
that the first term within brackets is regularization dependent:
we will see shortly that it is cancelled by a similar term from
another diagram.

The contribution of the photino-smuon diagrams of 
class (e) is in general divergent:
\bea
\label{divg}
\delta a_{\mu}^{(e)} & = & - \gamma_f {m_{\mu} \mgau \over 
16 \pi^2 \lsq} 
\left[  
2 \Delta_{UV} 
- \int_0^1 \! \! \! \! dx 
\left(
\log { m_1^2 x + \mgau^2 (1-x) - m_{\mu}^2 x(1-x) \over {\mu}^2}
+ (1 \to 2)
\right)
\right]
\nonumber
\\
& & -\gamma_f {m_{\mu}^2 {\delta}m^2  \over 8 \pi^2 \lsq} 
{ 1 \over m_1^2 - m_2^2}
\int_0^1 \! \! \! \! dx \, 
x \log { m_1^2 x + \mgau^2 (1-x) - m_{\mu}^2 x(1-x)
\over m_2^2 x + \mgau^2 (1-x) - m_{\mu}^2 x(1-x)} 
\, ,
\eea
where $\Delta_{UV}$ was defined in eq.~(\ref{deltauv}).
If $\gamma_f=0$, then $\delta a_{\mu}^{(e)} = 0$.

The diagrams of class (f), containing muon-muon-sgoldstino 
loops, give a finite result:
\bea
\delta a_{\mu}^{(f)} &  = & 
+ {m_{\mu}^2 \over 16 \pi^2} \left[ 
\left( {\delta m^2 \over F} + \gamma_K {F \over \lsq} \right)^2 
\int_0^1 dx {x^2 (2-x) \over m_S^2 (1-x) + m_{\mu}^2 x^2} \right.
\nn \\ & & \nn \\ & &
\;\;\;\;\;\;\;\;\; + \left.
\left( {\delta m^2 \over F} - \gamma_K {F \over \lsq} \right)^2 
\int_0^1 dx {- x^3 \over m_P^2 (1-x) + m_{\mu}^2 x^2} \right]
\, ,
\label{amb}
\eea

The overall contribution of the diagrams of class (g), 
containing the photon-muon-sgoldstino loops, is in general
divergent:
\bea
\label{dive}
\delta a_{\mu}^{(g)} &  = &  
\! \! \! {\gamma_K} {m_{\mu} \mgau \over 4 \pi^2 \lsq}
\left[ \Delta_{UV} - {1\over 2}
- \int_0^1 \! \! \! \! dx \int_0^{1-x} \! \! \! \! dy 
\left( 
\log { m_S^2 y + m_{\mu}^2 x^2 \over \mu^2}
\! + \! (S \to P)
\! + \! {m_{\mu}^2 x^2 \over m_P^2 y + m_{\mu}^2 x^2 } 
\right)
\right]
\nonumber
\\
&  & + {m_{\mu} \delta m^2 \mgau \over 8 \pi^2 F^2} \!
\left[ {\hat{\Delta}} -
2 \int_0^1 \! \! \! \! dx \int_0^{1-x} \! \! \! \! dy 
\left( \log { m_S^2 y + m_{\mu}^2 x^2 
\over m_P^2 y + m_{\mu}^2 x^2 }
- {m_{\mu}^2 x^2 \over m_P^2 y + m_{\mu}^2 x^2 } 
\right)
\right]
  \, ,
\eea
where $\Delta_{UV}$ and ${\hat{\Delta}}$ were defined in 
eqs.~(\ref{deltauv}) and (\ref{damureg}), respectively.
If $\gamma_K=0$, the result is finite. Notice also that
the regularization-dependent contribution in the second
line of eq.~(\ref{dive}) cancels exactly with the one of 
eq.~(\ref{amff}).

As a consistency check of our results, we have verified that $a_{\mu}$ 
vanishes in the limit of exact supersymmetry, as expected on general 
grounds \cite{fr}. We recall that, in the unbroken phase of 
supersymmetric QED, the overall $a_{\mu}$ is zero, as we can easily 
check from eq.~(\ref{amsqed}) by putting $m_1^2=m_2^2=m_{\mu}^2$ and 
$\mgau=0$:
\be
\label{sqedcan}
\delta a_{\mu}^{(b)} = - a_{\mu}^{(a)} 
= - {\alpha \over 2 \pi} \, .
\ee
Similar cancellations take place among the other diagrams.
Indeed, if we re-express the spectrum parameters in terms 
of the original ones and take the appropriate $F \to 0$ limit, 
we find:
\be
\delta a_{\mu}^{(f)}= -\delta a_{\mu}^{(c)}
={\rho^2 \over 16 \pi^2} \, , 
\;\;\;\; 
\delta a_{\mu}^{(g)}=-\delta a_{\mu}^{(d)}
={\eta \rho \, m \over 8 \pi^2 \Lambda} ({\hat{\Delta}}+1) 
\, , 
\;\;\;\; 
\delta a_{\mu}^{(e)}=0 \, .
\ee
Incidentally, this confirms that the cancellation of the terms 
$\pm m_{\mu} \delta m^2 \mgau {\hat{\Delta}}/(8 \pi^2 F^2)$ 
in the general results above is just a consequence of supersymmetry. 

\vspace{1cm}
\bc
{\bf 4.  Discussion}
\ec

We now discuss the interpretation of our results. We start from the
observation that, in an effective theory with spontaneously broken 
global supersymmetry and arbitrary defining functions $(w,K,f)$, the 
one-loop contribution to $a_{\mu}$ is in general divergent [see 
eqs.~(\ref{divg}) and (\ref{dive})]. This is in agreement with the 
results of ref.~\cite{paco}, obtained within the supergravity 
formalism. It is also consistent with the fact that we could have 
supplemented our standard (two-derivative) effective supersymmetric 
Lagrangian with higher derivative operators contributing directly to 
$a_{\mu}$. An example of such a `counterterm' is \cite{sugra}
\be
\label{superc}
{c \over \Lambda^3} \int d^4 \theta \, \left[
\ov{Z} W^{\alpha} M \stackrel{\longleftrightarrow}{D_{\alpha}} 
M^c + {\rm h.c.} \right] \, ,
\ee
where $W^{\alpha}$ is the supersymmetric field strength, $D_{\alpha}$ 
the covariant derivative of supersymmetry and $c$ an arbitrary 
coefficient. The expansion of (\ref{superc}) in component
fields contains the operator
\be
\label{magmom}
-{2 i c \over \Lambda^3} \, \ov{F}^{\ov{Z}} 
\mu \sigma^{\mu \nu} \mu^c F_{\mu \nu} + {\rm h.c.} \, ,
\ee
which in turn becomes a magnetic moment operator\footnote{ 
For convenience, in this section $F_{\mu \nu}$ denotes 
the original (not normalized) gauge field. Also, we use 
the definition $F^Z \equiv Z|_{\theta \theta}$, so that 
$\langle F^Z \rangle = - F$ in our model.} 
if supersymmetry is spontaneously broken by a non-vanishing 
VEV for the auxiliary field $F^Z$.
 
It is useful to reinterpret the divergent contributions to 
$a_{\mu}$ found in eqs.~(\ref{divg}) and (\ref{dive}) using 
a similar language, separating the perturbative generation of
supersymmetric operators from the question whether supersymmetry 
is broken or not. Thus we could say that our effective supersymmetric
Lagrangian {\em generates} an operator of the form (\ref{magmom}) 
at the one-loop level, with a logarithmically divergent coefficient
${\hat c}=\eta (\gamma_f-2\gamma_K) \Delta_{UV}/(32\pi^2)$.
We can go further and try to characterize such divergences in 
a more general way, manifestly invariant under analytic field 
redefinitions. To this purpose, we observe that ${\hat c}$ itself 
should be viewed as the VEV of a field dependent, reparametrization 
covariant expression, i.e. the operator should be written
as\footnote{The scalar field dependence of the fermionic terms that 
we are discussing is similar to that of certain bosonic terms of 
ref.~\cite{maryk}, where the logarithmically divergent bosonic 
contributions to the one-loop effective action of supergravity models 
were computed. It is plausible that our terms are supersymmetric
partners of some of those of ref.~\cite{maryk}, although we were 
unable to establish such a connection in full detail.}
\be
\label{divsusy}
-i {\Delta_{UV} \over 32 \pi^2} \, 
\langle { \ov{f}_{\ov{Z}} D_M D_{M^c} f \over ({\rm Re}f)^2 }
+ 2 {f_Z R^Z_{\; M \ov{Z} M^c} \over {\rm Re} f} \rangle \,
\ov{F}^{\ov{Z}} \,
\mu \sigma^{\mu \nu} \mu^c F_{\mu \nu} + {\rm h.c.} \, ,
\ee
where $D_i$ ($i = Z, M, M^c$) is the scalar field reparametrization 
covariant derivative and $R^{l}_{\; i \ov{n} j}$ the curvature 
tensor of the K\"ahler manifold:
\be
D_i D_j f = f_{ij} - f_l (K^{-1})^{l \ov{m}} K_{\ov{m} i j}
\;\; , \;\; R^{l}_{\; i \ov{n} j}=
(K^{-1})^{l \ov{m}} (K_{\ov{m} i \ov{n} j} - K_{\ov{m} \ov{n} p}
(K^{-1})^{p \ov{q}} K_{\ov{q} i j}) \, .
\ee 
For example, with our coordinate choice of eqs.~(\ref{w})--(\ref{k}), 
only the $\gamma_f M M^c$ term in $f$ and the $\gamma_K \ov{Z}^2 M M^c$ 
term in $K$ contribute to $\langle D_M D_{M^c} f \rangle$ and
$\langle R^Z_{\; M \ov{Z} M^c} \rangle$, respectively. We recall 
that other terms that could contribute to $\langle D_M D_{M^c} f 
\rangle$ and $\langle R^Z_{\; M \ov{Z} M^c} \rangle$ were removed by 
suitable field redefinitions. Our general expression (\ref{divsusy}) 
can be also used to reinterpret the divergent results of \cite{paco},
where no $M M^c$ term was included in $f$, but a $\ov{Z} M M^c$
term was present in $K$. We can view this as an alternative coordinate
choice, which leads to a non-vanishing $\langle D_M D_{M^c} f 
\rangle$ through the connection term [more specifically, 
from a type-(d) rather than a type-(e) diagram]. We should add that 
the supergravity calculation of \cite{paco} found several divergent 
contributions in individual diagrams, most of which were cancelling 
in the final result and two of which were not. In our case, the use 
of a globally supersymmetric Lagrangian (in a convenient field 
parametrization) avoids the proliferation of divergences and 
reproduces directly the two `genuine' ones.

Our covariant formula (\ref{divsusy}) clearly shows
the geometrical meaning of the two kind of divergences. 
We could now ask whether they should be considered
as independent or could possibly originate from a 
single object. In fact, taking into account that
$\langle f_M \rangle = \langle f_{M^c} \rangle=0$,
we observe that the VEV in eq.~(\ref{divsusy})
is identical to the VEV of $-4 D_{\ov{Z}} D_M D_{M^c} 
\log {\rm Re}f$. Thus we are led to conjecture that  
a (two-derivative) supersymmetric Lagrangian like
ours, with generic $f$ and $K$ functions, generates 
radiatively the following logarithmically divergent operator:
\be
\label{compact}
{i {\Delta_{UV} \over 8 \pi^2 } \,
( D_{\ov{Z}} D_M D_{M^c} \log {\rm Re}f ) 
\, \ov{F}^{\ov{Z}}}
\mu \sigma^{\mu \nu} \mu^c F_{\mu \nu} + {\rm h.c.} \, .
\ee
The corresponding (higher-derivative) superfield operator
should be an appropriate generalization of the operator 
(\ref{superc}).

We have discussed so far the structure of the supersymmetric
operators that give divergent contributions to $a_{\mu}$
once supersymmetry is broken. Since the existence of such 
operators limits somewhat our predictive power, we could
look for models where the corresponding coefficients 
vanish, or are at least suppressed. The simplest case 
\cite{sugra,mo,paco} is to have $\langle f_Z \rangle =
0$, implying $\mgau=0$. More interestingly, the whole 
combination in eq.~(\ref{divsusy}) [or the corresponding 
compact expression in eq.~(\ref{compact})] might vanish if 
$f$ and $K$ were related, either functionally or just on the 
vacuum (the latter case corresponds to the equality $\gamma_f
=2\gamma_K$, if we use our original parametrization). This approach 
may lead to models with an underlying extended supersymmetry. 

The question could also be addressed in terms of ordinary symmetry 
arguments. We may ask whether there is a symmetry that allows for 
non-vanishing values of $m_{\mu}$ and $a_{\mu}$ but forbids the 
divergent one-loop contributions. A possible candidate is the 
R-symmetry defined by the following charge assignments: $R(Z)=+2$, 
$R(M M^c)=+2$ (the separate
values of the R-charges of $M$ and $M^c$ are not important here).
Such an assignment, however, would put to zero not only the 
coefficients $\gamma_f$ and $\gamma_K$ of the divergent operators
(and possible coefficients related to them by analytic field 
redefinitions compatible with the R-symmetry), but also the mixing 
terms in the smuon  and sgoldstino mass matrices and, most importantly,
the photino mass $\mgau$. With the latter constraint, the construction
of a realistic model with the full Standard Model gauge group appears
very unlikely. We were unable to find alternative symmetries that 
forbid the divergent one-loop contributions to $a_{\mu}$ in the
presence of a realistic spectrum.

A milder requirement may be to ask for a symmetry that forces the 
divergent contributions, eqs.~(\ref{divg}) and (\ref{dive}), to 
be at least proportional to $m_{\mu}^2$. If so, reasonable choices
of the ultraviolet cutoff would produce contributions to $a_{\mu}$
that are at most of the same order of magnitude as the finite ones.
An obvious candidate for such a symmetry is a chiral $U(1)_S$,
with the charge assignments: $S(Z)=0$, $S(M M^c) \ne 0$. Such a symmetry
would be explicitly broken by a small parameter $\epsilon$ to allow 
for the muon mass term. As a consequence, also $\delta m^2$, $\gamma_f$ 
and $\gamma_K$ would be suppressed by the same small parameter. 
In particular, we would obtain $\delta m^2 \sim m_{\mu} A$, with $A$ 
of order the other supersymmetry-breaking masses. Such a hierarchy 
would be stable against radiative corrections, as can be easily 
checked by looking at the one-loop diagrams contributing
to $m_{\mu}$ and to $\delta m^2$. 

In order to discuss in more detail the relative size of 
the ${\cal{O}} (m_{\mu}^2)$ contributions to $a_{\mu}$, we 
should supplement the chiral power counting with additional 
information. For example, we could assume\footnote{We thank
the referee for raising this point.} that $F/\Lambda^2 \sim 
\epsilon'$ is small and consider an expansion in both $\epsilon$ 
and $\epsilon'$. Even so, at least two possibilities could be 
considered for the combined power counting: (i) $ \gamma_K \sim 
\gamma_f \sim \epsilon$, $m_{\mu} \sim \epsilon \Lambda \,$; 
(ii) $ \gamma_K \sim \gamma_f \sim \epsilon$, $m_{\mu} \sim 
\epsilon \, \epsilon' \Lambda \, $.
In case (i), the contributions to  $a_{\mu}$ proportional
to $\gamma_K$ and $\gamma_f$ [or induced by the counterterm
of eq.~(\ref{superc})] would be suppressed by a factor 
$\epsilon'$ with respect to the remaining ones, and could
be considered as subleading. Then we would obtain an unambiguous 
result at the leading order. In case (ii), on the other hand, 
such contributions would be formally of the same order of the 
other ones\footnote{Notice that the doubly suppressed muon mass 
of case (ii) would naturally arise from a K\"ahler potential
term of the form $\beta_K \ov{Z} M M^c/\Lambda$, with 
$\beta_K \sim \epsilon$. More generally, case (ii) can be 
justified by an R-symmetry under which $Z, M, M^c$ are neutral,
explicitly broken by the parameter $\epsilon'$: then all 
superpotential parameters would be suppressed by $\epsilon'$. 
Incidentally, this would also imply the same scaling law for 
all sgoldstino mass terms, $(m_S^2 \pm m_P^2) \sim F^2/\Lambda^2$.}.
Moreover, as we shall discuss in more detail below, phenomenological 
considerations will lead us to consider, among others, values of $F /
\Lambda^2$ of order one.

We can summarize the above discussion as follows. The effective 
supersymmetric low-energy theory can contain higher-derivative 
terms that contribute to $a_{\mu}$ and do not depend (only) on 
the spectrum. Such terms could either be present as counterterms 
since the beginning, or be generated radiatively within the 
low-energy theory itself, or both. Symmetry considerations could 
help in keeping the size of such contributions under control, 
as in the example of a chiral symmetry, possibly combined with 
an expansion in $F/\Lambda^2$. In most phenomenologically relevant 
situations, however, it seems that only a better knowledge of the 
underlying microscopic theory could remove the residual ambiguities.  
Taking into account all this, in the following we will use the finite 
contributions to $a_{\mu}$ only to derive some `naturalness' 
constraints on the model parameters, barring the possibility 
of accidental (or miraculous) cancellations. 

\newpage 

\vspace{1cm}
\bc
{\bf 5. Phenomenology}
\ec

Keeping in mind the discussion of the previous section, we now 
focus on the class of models in which 
\be
\label{finite}
\langle D_M D_{M^c} f \rangle = 0 \;\;\ , \;\;
\langle R^Z_{\; M \ov{Z} M^c} \rangle = 0 \, ,
\ee
so that the total one-loop contribution to $a_{\mu}$ is 
finite and determined only by the parameters controlling the 
spectrum\footnote{The finiteness conditions (\ref{finite}) 
generalize those of ref.~\cite{paco}. In our original 
parametrization, they amount to set $\gamma_f=\gamma_K=0$. 
We will not discuss here the alternative
intriguing possibility mentioned in the previous section, 
corresponding to $\gamma_f = 2 \gamma_K \ne 0$. 
We only recall that also in the latter case the result would be finite
and regularization-independent, as long as we use regularizations 
compatible with supersymmetry. However, it would contain an additional 
parameter besides those controlling the spectrum.}. 
We specialize the general formulae of sect.~3 to some 
interesting limits, and we discuss the resulting phenomenological 
constraints. 

We begin with the SQED contribution. For realistic
smuon and photino masses, so that the muon mass becomes negligible 
in the comparison, eq.~(\ref{amsqed}) becomes
\bea
\delta a_{\mu}^{(b)} & = & - {\alpha \over 2 \pi} \, m_{\mu}^2 \,
\left[ {m_1^6 - 6 m_1^4 \mgau^2 + 3 m_1^2 
\mgau^4 + 2 \mgau^6 + 6 m_1^2 \mgau^4 \log (m_1^2/\mgau^2)
\over 6 ( m_1^2 - \mgau^2)^4} 
+ (1 \to 2) \right] \nn \\ & & \nn \\ & & + 
{\alpha \over \pi}  {m_{\mu}  \, \delta m^2 \, \mgau 
\over m_1^2 - m_2^2} \left[ {m_1^4 -  \mgau^4 - 2 m_1^2 
\mgau^2 \log (m_1^2/\mgau^2) \over 2 ( m_1^2 - \mgau^2)^3} 
- (1 \to 2) \right]  \, .
\label{amlsqed}
\eea

Next, we consider the contributions from the smuon-photino-goldstino
sector, for $\gamma_f=0$. Eq.~(\ref{divg}) gives no contribution,
whereas eqs.~(\ref{ama}) and (\ref{amff}) simplify considerably in 
the phenomenologically relevant limit $m_{\mu} \ll m_1,m_2,\mgau$:
\be
\label{amaa}
\delta a_{\mu}^{(c)} = - {m_{\mu}^2 (m_1^2 + m_2^2) 
\over 96 \pi^2 F^2} \, ,
\ee
\be   
\label{amf}
\delta a_{\mu}^{(d)} =  
- {m_{\mu} \delta m^2 \mgau \over 8 \pi^2 F^2} {\hat{\Delta}}
+ {m_{\mu}^2 \over 16 \pi^2 F^2} 
\left[ {m_1^2 \mgau^2 \over m_1^2 - \mgau^2} 
\left( 1- {\mgau^2 \over m_1^2 - \mgau^2} \log {m_1^2
\over \mgau^2} \right) + (1 \to 2) \right] \, .
\ee
The result in eq.~(\ref{amaa}) is in full agreement with 
ref.~\cite{bm}, and disagrees in sign with the supergravity 
computations of refs.~\cite{mo,paco}. The result in 
eq.~(\ref{amf}) can be compared with a similar expression
of ref.~\cite{paco}, where the terms within brackets
appear with opposite sign. Also, we recall that the first term 
of eq.~(\ref{amf}) is regularization-dependent and disappears 
when all contributions are summed.

We now discuss the contributions from the muon-photon-sgoldstino
sector, eqs.~(\ref{amb}) and (\ref{dive}), for $\gamma_K=0$.
Before taking any kinematical limit, we would like to emphasize 
that such contributions depend only on the spectrum parameters 
and the supersymmetry breaking scale, with no more model-dependence 
than the contributions from the smuon-photino-goldstino sector. 
This point was apparently overlooked in ref.~\cite{paco}.
However, we have checked that the results for the muon-photon-sgoldstino 
sector given there agree with ours, once the flat limit is 
taken and appropriate parameter identifications 
are made [e.g., $3 m_{3/2}(m_{\mu}-m'')=\delta m^2$].
Furthermore, the general results in eqs.~(\ref{amb}) and 
(\ref{dive}) simplify considerably in the limit of
heavy or light sgoldstinos:
\be
\delta a_{\mu}^{(f)} = + {(\delta m^2)^2  \over 16 \pi^2 F^2}
\times \left\{ \begin{array}{ll}
{m_{\mu}^2 \over m_S^2} \left( - {7 \over 6} +
\log {m_S^2 \over m_{\mu}^2} \right) + {m_{\mu}^2 \over m_P^2} 
\left( {11 \over 6} - \log {m_P^2 \over m_{\mu}^2} \right)
& (m_{\mu} \ll m_P,m_S)
\\ & \\
1
&
(m_S,m_P \ll m_{\mu})
\end{array}
\right.
\, ,
\label{ambbb}
\ee
\be
\delta a_{\mu}^{(g)} = 
+ {m_{\mu} \delta m^2 \mgau \over 8 \pi^2 F^2} 
{\hat{\Delta}}
+ {m_{\mu} \delta m^2 \mgau \over 8 \pi^2 F^2} 
\times \left\{ \begin{array}{ll}
\log {m_P^2 \over m_S^2} 
& 
(m_{\mu} \ll m_P,m_S)
\\ & \\
1
& 
(m_P,m_S \ll m_{\mu})
\end{array}
\right.
\, .
\label{ambe}
\ee
Taking other special limits such as $m_{P(S)} \ll m_{\mu} \ll m_{S(P)}$
is equally straightforward.
We recall that the possibility of superlight sgoldstinos (both of 
them or just one) has been frequently considered in the superlight 
gravitino literature \cite{lsg}. Despite its naturalness problems
\cite{bfzff}, it may be related with possible dynamical mechanisms 
for the resolution of the strong-CP and/or the cosmological constant 
problems.

We conclude this section by confronting the above results with
the present and future experimental constraints\footnote{For
previous analyses based on the one-loop goldstino contributions
of ref.~\cite{paco}, see refs.~\cite{lln} and \cite{fergri}. The 
latter paper studies also the influence of superlight sgoldstinos, 
but neglects their effects at the one-loop level and only 
partially includes them at the two-loop level.}. In the following, 
it will be convenient to parametrize the off-diagonal element 
of the smuon mass matrix as $\delta m^2 \equiv m_{\mu} A$,
as suggested for example by an underlying approximate chiral 
symmetry (see sect.~4). Thus, although $A$, as defined by the 
previous relation, is a free parameter equivalent to 
$\delta m^2$, it is natural to expect that $A$ be not
much larger than the other supersymmetry breaking masses.
We begin by reconsidering the SQED contribution of 
eq.~(\ref{amlsqed}). We recall that such a contribution
should be replaced, in a fully realistic model, by the full 
MSSM contribution \cite{gm2mssm}: here it will be used only 
as a benchmark for the new contributions associated 
with the superlight gravitino and its superpartners.
For illustration purposes, we can further simplify 
eq.~(\ref{amlsqed}) and consider just a representative 
case. For example, if the smuons and the photino are 
roughly degenerate, with a common mass $M_S$, we obtain
\be 
\label{amsqedbis}
\delta a_{\mu}^{(b)} = - {\alpha \over 12 \pi} 
{m_{\mu}^2 \over M_S^2} \left( 1 + {A 
\over M_S} \right) \simeq -2 \times 10^{-10}
\times {(100 \, {\rm GeV})^2 \over M_S^2} 
\times \left( 1 + {A \over M_S} \right)
\, . 
\ee  
As we can see, sizeable contributions can be obtained only if the 
sparticle masses are significantly smaller than 100~GeV or 
if the supersymmetry-breaking mass parameter $A$ is significantly 
larger than the typical supersymmetry-breaking 
masses\footnote{We recall that, in the framework of the MSSM,
the role of our effective $A$ parameter would be played 
by the MSSM parameters $A_{\mu}$ and $\mu\tan\beta$.
Thus, for example, enhancement effects could originate 
from large values of $\tan\beta$ \cite{gm2mssm}.}.

\begin{figure}[ht]
\vspace{-0.01cm}
\epsfig{figure=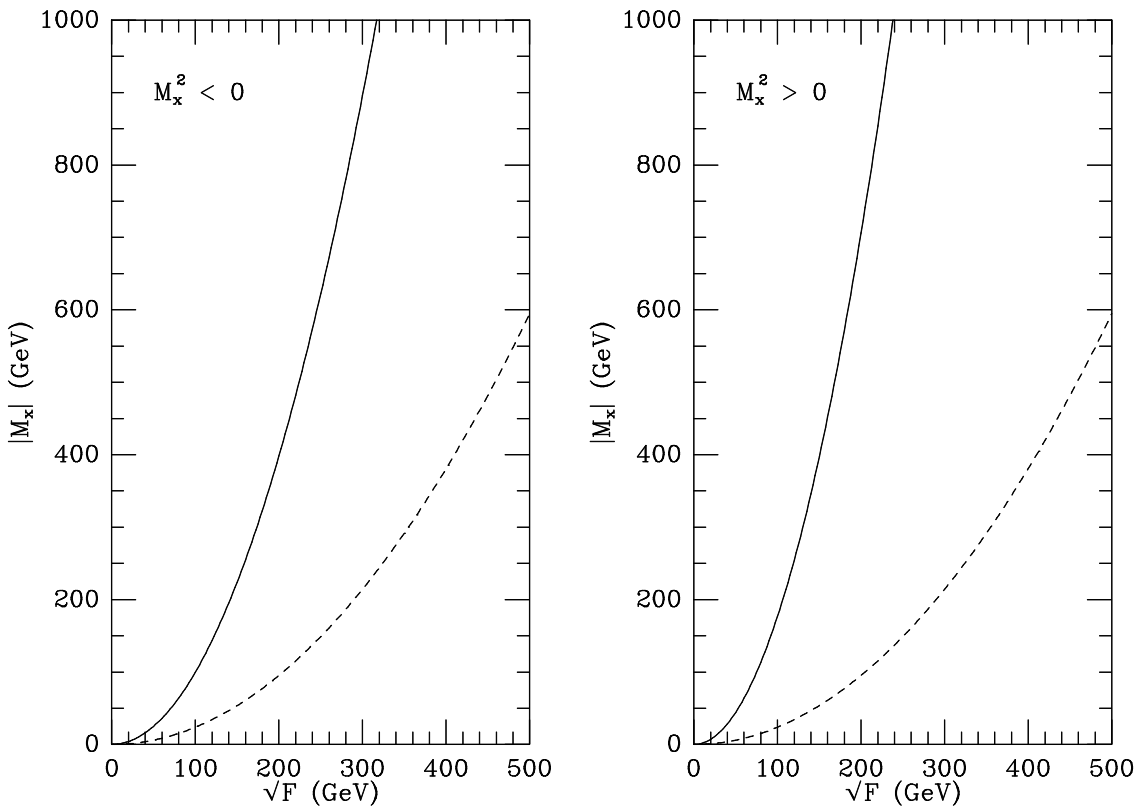,height=6cm,angle=-90}
\vspace*{3.0cm}
\caption{{\it Contours of $\delta a_{\mu} \equiv m_{\mu}^2
M_x^2 / (16 \pi^2 F^2)$ in the plane $(\sqrt{F},|M_x|)$: the 
regions above the solid lines correspond to $\delta a_{\mu} 
< - 70 \; (> + 220) \times 10^{-10}$, and the dashed lines 
correspond to $\delta a_{\mu} = \mp 4 \times 10^{-10}$.}}
\label{ccc}
\end{figure}

To discuss the phenomenological impact of the remaining contributions,
it will be useful to parametrize all of them in a uniform fashion
\be
\label{param}
\delta a_{\mu} \equiv {m_{\mu}^2 M_x^2 \over 16 \pi^2 F^2} \, ,
\ee
where the real parameter $M_x^2$ has the dimension of a mass squared
and can be positive or negative. The above form separates the 
dependence of $\delta a_{\mu}$ on the spectrum (through $M_x$) 
from that on the supersymmetry-breaking scale $\sqrt{F}$. 
Incidentally, we recall that the latter scale is related to 
the gravitino mass $m_{3/2}$, the reduced Planck mass $M_P$ 
and Newton's constant $G_N$ as follows: 
$F^2 = 3 m_{3/2}^2 M_P^2 = 3 m_{3/2}^2/( 8 \pi G_N)$.
The present experimental limit and the future sensitivity on $M_x$ 
are then shown in fig.~\ref{ccc}: the two solid lines correspond
to the present upper and lower bounds on $\delta a_{\mu}$, see
eq.~(\ref{limit}), whereas the dashed line corresponds to $|\delta
a_{\mu}| = 4 \times 10^{-10}$, and gives a measure of the expected
future sensitivity. 

The information in fig.~2 should be combined with 
the explicit expression of $M_x^2$ in terms of the 
spectrum, which can be easily read off our results.
In particular, we can put together our results of 
eqs.~(\ref{amaa})--(\ref{ambe}) and express $M_x^2$ 
as the sum of `goldstino' and `sgoldstino' contributions,
\be
\label{deco}
M_x^2 = M^2_{x(G)}+M^2_{x(SG)} \, ,
\ee
where 
\be
\label{mxg}
M^2_{x(G)} = m_1^2 \left[ - {1 \over 6} + 
{\mgau^2 \over m_1^2 - \mgau^2} 
\left( 1- {\mgau^2 \over m_1^2 - \mgau^2} \log {m_1^2
\over \mgau^2} \right) \right] + (1 \to 2) \, ,
\ee
and 
\bea 
\label{mxsg}
M^2_{x(SG)} =
\left\{ \begin{array}{ll}
2 A \mgau \log {m_P^2 \over m_S^2} 
+ {\cal O}(A^2 m_{\mu}^2/m_{S,P}^2)
& (m_{\mu} \ll m_P,m_S)
\\ & \\
2 A \mgau  + A^2
&
(m_S,m_P \ll m_{\mu})
\end{array}
\right.
\, .
\eea
Notice that we have already removed from each contribution
the regularization-dependent terms that disappear 
in the overall result.

The size of the goldstino contribution to $a_{\mu}$ depends 
mainly on the smuon masses, and can be significant only for 
heavy smuons and a very low supersymmetry-breaking scale. 
The corresponding sign changes from positive to negative
for increasing smuon-to-photino mass ratio. If the smuons 
and the photino have roughly the same mass $M_S$, we obtain 
$M^2_{x(G)} \simeq 2 M_S^2/3$. In interpreting fig.~\ref{ccc},
however, we should keep in mind that the situations where 
$|M_{x(G)}| \gg \sqrt{F}$ correspond to a strong-coupling 
regime, and are limited by perturbative unitarity. As discussed 
for example in \cite{bfz3,bfzff}, a reasonable requirement is 
$|M_{x(G)}| \simlt (2 \div 3) \sqrt{F}$.

The size of the sgoldstino contribution to $a_{\mu}$
depends on the photino mass, the sgoldstino masses
and the parameter $A$. The latter parameter, which 
controls the off-diagonal element in the smuon mass 
matrix, plays a crucial role. In particular, 
the sgoldstino contribution is zero for vanishing $A$.
Several possibilities can arise for non-vanishing $A$.
We comment here on some representative cases,
using the symbol $M_S$ to denote a typical smuon or
photino mass. Intermediate cases not discussed below 
can be easily worked out from our general results.
\begin{description}
\item[i)]
{\em Heavy sgoldstinos.} By `heavy' here we mean the 
case $m_S,m_P = {\cal O}(M_S)$, which is favoured
by naturalness considerations \cite{bfzff}.  
Then from eq.~(\ref{mxsg}) we have $M^2_{x(SG)} 
\simeq 2 A\mgau \log(m_P^2/m_S^2)$,
with either sign being possible. Notice that
the sgoldstino contribution becomes negligible 
if $S$ and $P$ are approximately degenerate. 
Otherwise, we should also specify the size of $A$.
In the natural case of $A={\cal O}(M_S)$, 
$M^2_{x(SG)}$ is ${\cal O}(M_S^2)$ as well, thus
the sgoldstino contribution to $a_{\mu}$
is ${\cal O}(m_{\mu}^2 M_S^2/16 \pi^2 F^2)$,
like the goldstino one. However, if (for some reason) 
$A \gg M_S$, the sgoldstino contribution gets enhanced 
by a factor ${\cal O}(A/M_S)$ compared to the goldstino one 
[for a similar situation in SQED, see eq.~(\ref{amsqedbis})].
\item[ii)] {\em Superlight sgoldstinos.} By `superlight'
we mean here the case $m_S, m_P \ll m_{\mu}$ (however,
the qualitative picture remains the same whenever at 
least one of the sgoldstino masses is $\simlt m_{\mu}$).
Then from eq.~(\ref{mxsg}) we have $M^2_{x(SG)} =  2 A
\mgau + A^2 $. Again, if $A={\cal O}(M_S)$, then 
$M^2_{x(SG)}$ is ${\cal O}(M_S^2)$ (either sign being 
possible), and the sgoldstino contribution is comparable
with the goldstino one. If instead $A \gg M_S$, then we 
have $M^2_{x(SG)} \simeq A^2 \gg M^2_{x(G)}$, and the
sgoldstino contribution to $a_{\mu}$ 
($=m_{\mu}^2 A^2/16 \pi^2 F^2$) dominates over the 
goldstino one. Thus the simultaneous occurrence 
of superlight sgoldstinos, large $A$ 
parameter and low supersymmetry-breaking scale
gives a potentially large contribution to $a_{\mu}$, 
comparable with (or even exceeding) the present 
experimental bounds. However, it should be taken 
into account that such a situation is an extreme one, 
as we have already remarked. Moreover, other
potential enhancement effects due to superlight 
sgoldstinos have been studied in the context of 
cosmology, astrophysics or collider physics 
\cite{lsg}, and the inferred lower bounds on 
$\sqrt{F}$ are already well above the TeV range,
weakening the impact of the $a_{\mu}$ constraint.
\end{description}

In summary, since the present lower limits from accelerator searches 
should be of the order of 100 GeV for the smuon and photino 
masses\footnote{A dedicated analysis for models with a superlight 
gravitino is not available yet, and would require the generalization 
of the model to include the full SM gauge group.}, and of the order 
of 200~GeV for $\sqrt{F}$ \cite{bfzph}, we can see that at present 
$a_{\mu}$ provides a non-negligible but mild indirect constraint. 
Such a constraint, however, will become much more stringent after 
the completion of the E821 experiment. Should a discrepancy emerge
between the future E821 result and the SM prediction, models with 
a superlight gravitino may provide a viable explanation.

\vspace{1cm}
\bc
{\bf 6. Final comments}
\ec

We conclude this work with some comments on the predictive power of 
our effective-theory approach, to clarify its advantages with respect 
to non-linear realizations of supersymmetry, as well as its residual 
limitations.

Since $a_{\mu}$ is a low-energy observable involving only SM particles, 
we may na\"{\i}vely think that its calculation could be addressed in 
the context of a `more effective' theory, where heavy particles have 
been integrated out and supersymmetry is non-linearly realized. In the 
case of heavy sgoldstinos, a first step in this direction would be to 
integrate out only the sgoldstino fields, ending up with a non-linear 
goldstino superfield \cite{bf}. A more radical step would be to integrate 
out not only the sgoldstinos but also the smuons and the photino, ending
up with a fully non-linear realization of supersymmetry \cite{nonlin}
on the muon, photon and goldstino fields. It is immediate to realize,
however, that non-linear realizations are not a sufficiently predictive 
approach to address the computation of $a_{\mu}$. This should be 
contrasted with other processes involving external goldstinos 
\cite{bfz3,bfzph}, where the low-energy theorems of spontaneously 
broken supersymmetry can be put to work. The simplest way to see this 
point is to look at the dependence of our general finite result 
(\ref{param}) on the supersymmetry-breaking masses, rewriting it as:
\be
\label{nuova}
\delta a_{\mu} = {m_{\mu}^2 M_S^2 \over F^2} \; G \! 
\left( {m_i^2 \over M_S^2} \right) \, ,
\ee
where $M_S$ is a typical supersymmetry-breaking mass that can 
be taken as a reference scale, the $m_i^2$ are the individual 
supersymmetry-breaking mass parameters in the sgoldstino, smuon 
and photino sectors, and the dimensionless function $G$ encodes 
detailed information on the supersymmetry-breaking spectrum. From 
the explicit form of $G$ [see, e.g., eqs.~(\ref{deco})--(\ref{mxsg})], 
it is evident that there are no decoupling properties for particles 
with large supersymmetry-breaking masses. This result is expected:
decoupling should not occur for heavy particles whose masses break 
a symmetry that would set the result to zero. In the case of a 
non-linear realization, we could at most obtain the general scaling 
law of eq.~(\ref{nuova}), but the function $G$ would be just replaced 
by a free parameter (if we replace the linear superfield $Z$ by a 
non-linear goldstino superfield, also the form and the dimension of 
the lowest-dimensional counterterm for $a_{\mu}$ will change). 
Starting from an effective theory where some of the superpartners are 
not present would in general reduce our predictive power, generating 
new divergences without a prescription on how to regulate them. In 
this respect, removing only the sgoldstinos, while keeping the smuons 
and the photino, would not lead to any substantial increase in 
predictive power: sgoldstinos do not have any special decoupling 
properties with respect to other particles that may acquire large
supersymmetry-breaking masses.

The second and last comment is a warning on the correct interpretation
of the phenomenological implications of our results. As already discussed
in sect.~4, the finite result of our general calculation, obtained by
setting $\gamma_K = \gamma_f =0$, is protected against large corrections
only when $\sqrt{F} \ll \Lambda$ and there is a symmetry guaranteeing that
$\gamma_K \sim \gamma_f \sim m_{\mu} / \Lambda$. In this case, the only
realistic possibility of generating observable contributions to $\delta 
a_{\mu}$ is to have $A$ much larger than the other supersymmetry-breaking 
masses, to enhance the sgoldstino contributions of eq.~(\ref{mxsg}).
This is certainly possible, and is the exact counterpart of having 
large values for $A_{\mu}$ and/or $\mu \, \tan \beta$ in the MSSM.
However, the value of $A$ cannot be pushed too far, since $A$ breaks
the same chiral symmetry that protects the smallness of $m_{\mu}$:
too large values for $A$ would make $m_{\mu}$ unnaturally small.
Apart from the above possibility, observable contributions to 
$\delta a_{\mu}$ can be obtained only when $\gamma_K$ and $\gamma_f$
are unsuppressed, in which case we have no predictive power at all,
or when $\sqrt{F} \sim \Lambda$. In the latter case, we can still
make use of our finite result, but we should keep in mind that we 
are neglecting other contributions of comparable size. If no deviations
from the SM predictions are found, we can still use the results to derive
some `naturalness' constraints on the model parameters, barring the 
possibility of accidental (or miraculous) cancellations. Should a
discrepancy emerge between the experimental data and the SM prediction, 
however, a detailed quantitative analysis would require a better 
knowledge of the underlying microscopic theory, to remove the
ambiguities associated with the operators controlled by $\gamma_K$
and $\gamma_f$ and with possible unsuppressed higher-derivative
supersymmetric operators. 

\vfill{
{\bf Acknowledgements. }
We would like to thank Ferruccio Feruglio and Stefano
Rigolin for useful discussions and suggestions.}

\newpage

\end{document}